\documentclass[twocolumn]{aastex701}
\usepackage{graphicx}        
\usepackage{amssymb}        
\usepackage{color}           
\usepackage{booktabs}
\usepackage{amsmath}
\usepackage{lineno}
\definecolor{altgreen}{RGB}{46, 139, 87}

\chardef\us=`\_

\graphicspath{{./}{figures/}}

\shorttitle{ Neupert Modelling}
\shortauthors{Tadepalli et al.}

\begin{document}

\title{Investigating the Neupert Effect with HEL1OS/Aditya-L1 }

\correspondingauthor{Srikar Paavan Tadepalli}
\email{srikar@ursc.gov.in}

\author{Srikar Paavan Tadepalli}
\affiliation{Space Astronomy Group, ISITE Campus, U R Rao Satellite Centre, Bengaluru, India}
\affil{Department of Physics, Indian Institute of Technology, Varanasi, India}
\email{srikar@ursc.gov.in}

\author{Abhishekh Kumar Srivastava}
\affil{Department of Physics, Indian Institute of Technology, Varanasi, India}
\email{asrivastava.app@iitbhu.ac.in}

\author{K. Sankarasubramanian}
\affiliation{Space Astronomy Group, ISITE Campus, U R Rao Satellite Centre, Bengaluru, India}
\email{sankark@ursc.gov.in}

\author{Manju Sudhakar}
\affiliation{Space Astronomy Group, ISITE Campus, U R Rao Satellite Centre, Bengaluru, India}
\email{manjus@ursc.gov.in}

\author{Anand Jain}
\affiliation{Space Astronomy Group, ISITE Campus, U R Rao Satellite Centre, Bengaluru, India}
\email{anandj@ursc.gov.in}

\author{Brajpal Singh}
\affiliation{Space Astronomy Group, ISITE Campus, U R Rao Satellite Centre, Bengaluru, India}
\email{brajpal@ursc.gov.in}

\author{Reenu Palawat}
\affiliation{Space Astronomy Group, ISITE Campus, U R Rao Satellite Centre, Bengaluru, India}
\email{reenu@ursc.gov.in}

\author{Ravishankar B.T.}
\affiliation{Space Astronomy Group, ISITE Campus, U R Rao Satellite Centre, Bengaluru, India} 
\email{ravibt@ursc.gov.in}

\author{Anuj Nandi}
\affiliation{Space Astronomy Group, ISITE Campus, U R Rao Satellite Centre, Bengaluru, India}
\email{anuj@ursc.gov.in}

\begin{abstract}
 We investigate the Neupert effect in solar flares using broadband hard X-ray observations from the HEL1OS instrument onboard Aditya-L1. By separating thermal and non-thermal components from HEL1OS spectrograms, we analyse a sample of 312 flares observed between November 2023 and May 2026. We find that the soft X-ray flux is strongly correlated with the time-integrated hard X-ray emission, with a median correlation coefficient of 0.984, indicating that Neupert-like behaviour is nearly ubiquitous in flares with detectable non-thermal emission. Extending beyond statistical correlation, we develop an empirical forward model in which the thermal lightcurve is reconstructed from the non-thermal emission by resolving the hard X-ray lightcurve into individual energy-release episodes and convolving their fitted profiles with finite thermal-response functions. This approach successfully reproduces the observed thermal evolution, including flares that exhibit quasi-periodic pulsations (QPPs). Our results demonstrate that the broadband HEL1OS observations sample both thermal-dominated and non-thermal-dominated flare emission within a single instrument, providing an observation-driven link between the temporal structure of non-thermal HXR emission and the thermal X-ray response. This provides a different framework for interpreting flare energetics and temporal evolution using broadband hard X-ray measurements.
\end{abstract}

\section{Introduction}
\label{S-intro}

The Neupert effect provides one of the most direct observational links between particle acceleration and plasma heating during solar flares \citep{1968ApJ...153L..59N}. In its classical form, the time-integrated microwave or hard X-ray (HXR) emission follows the evolution of the soft X-ray (SXR) flux \citep{1968ApJ...153L..59N,Dennis1993-wq}. This behaviour is naturally interpreted within the collisional thick-target model: electrons accelerated during the impulsive phase propagate towards the lower atmosphere, where they lose energy through Coulomb collisions and produce bremsstrahlung HXR emission. The associated energy deposition heats the chromospheric plasma and drives evaporation into coronal loops, producing the subsequent
thermal X-ray emission \citep{Brown1971-jh,2011SSRv..159..107H}. The Neupert relationship therefore connects two fundamental components of flare energetics---the non-thermal energy release and the response of the heated plasma.

The relationship is nevertheless not expected to be exact. The thermal emission depends not only on the instantaneous energy input but also on the evolving temperature and density of the plasma, conductive and radiative losses, and any heating that persists beyond the impulsive phase. Moreover, the conversion between observed HXR emission and the energy deposited by accelerated electrons depends sensitively on the low-energy part of the
electron distribution \citep{2011SSRv..159..107H,Kontar2015-zq}. Departures from ideal Neupert behaviour can consequently contain information on energy transport, continued heating and plasma cooling rather than simply representing a failure of the relationship.

Statistical studies have established that Neupert-like behaviour is common,
particularly among flares with detectable HXR emission. Using BATSE and GOES
observations, \citet{Veronig2002-ms} found the expected timing relationship in a
large fraction of their flare sample, while also identifying events in which
the SXR evolution continued beyond the principal HXR phase. More recently,
\citet{2024SoPh..299...57L} reported strong correlations between cumulative HXR emission from ASO-S/HXI and GOES SXR flux for 149 HXR-selected flares. Such studies demonstrate the prevalence of the effect, but the conventional correlation compresses the non-thermal emission into its cumulative fluence and, therefore, does not retain the temporal organisation of the underlying energy release.
That temporal information is important because flare HXR emission commonly consists of multiple impulsive bursts rather than a single heating episode~\citep{McAteer2007-sr,McAteer2013-nf,Dennis1985-in}.

The thermal lightcurve represents the combined response to these episodes, modified by the finite thermodynamic response of the plasma. In the limiting case where energy losses are negligible during the heating phase, this leads to the familiar integral Neupert relation; once cooling and continued heating become important, the thermal evolution need not be a simple integral of the HXR emission. Hydrodynamic models address this problem by evolving the plasma response to a prescribed heating function
\citep{1980SoPh...68..351N,1989SoPh..121..289P,2003A&A...401..699B, 2008ApJ...682.1351K,2012ApJ...752..161C,Allred2015}, but determining that heating function observationally remains a key difficulty.

Empirical approaches using convolution-based descriptions of the evolution of SXR flare have previously been explored. \citet{Aschwanden2012-cn} represented the SXR response as
the convolution of an energy-release function with exponential cooling,
while \citet{2017SoPh..292...77G} modelled elementary SXR flare profiles using
Gaussian energy release convolved with exponential decay. More recently,
\citet{Qiu2021-yk} used spatially resolved UV footpoint emission as an
observational proxy for heating individual flare loops and reproduced the
integrated SXR evolution using empirical cooling functions. HXR emission offers a complementary starting point: it directly traces the accelerated-electron population associated with impulsive energy release, while retaining its temporal substructure. A pulse-resolved HXR driver could therefore connect the classical Neupert relationship more
directly to the sequence of energy-release episodes that produces the
observed thermal evolution.

In this work, we use HEL1OS \citep{Nandi2025-ib} observations to address the Neupert relationship at two complementary levels. First, we test its prevalence and energy dependence across a large sample of flares with measurable non-thermal emission. We then ask whether the time-dependent thermal X-ray lightcurve can be reconstructed from the temporally resolved HXR driver itself. The resulting framework also provides a means of identifying phases in which the measured non-thermal emission is insufficient to account for the observed thermal response.

\section{Observations}
\label{S-observation}

The dataset used in this study is obtained from the HEL1OS instrument onboard Aditya-L1 ~\citep{Parate2025-mo,Sankarasubramanian2025}. HEL1OS is a Sun-as-a-star hard X-ray spectrometer operating in the energy range of 7–150 keV, designed to study solar flare emission over a broad spectral range. The instrument employs two detector systems -- Cadmium Telluride (CdTe) and Cadmium Zinc Telluride (CZT) -- to achieve high sensitivity across low and high X-ray energies, respectively. HEL1OS provides both event list and spectral data products as part of its Level 1 dataset ~\citep{Ravishankar2026}. The event list has a native timing resolution of 10 ms; however, for the purposes of this study, spectrograms are constructed at a cadence of 5 s to ensure adequate counting statistics across all energy bands. These spectrograms are corrected for instrumental effects, including dead time and temperature variations, and are subsequently converted into band-resolved lightcurves. To investigate the Neupert effect in an energy-resolved manner, the spectrograms are divided into multiple energy bands spanning both thermal- and non-thermal-dominated regimes. The CdTe detector bands are defined as 7–10 keV, 10–15 keV, 15–25 keV, and 25–40 keV, while the CZT detector bands are defined as 25–40 keV, 40–55 keV, and 55–70 keV. The master flare list contains $\sim456$ events observed between November
2023 and May 2026. After applying the selection criteria described in
Section~\ref{S-dataselect}, 312 flares are retained for the present analysis.


\section{Data analysis}
\label{S-analysis}

\subsection{Data Selection}
\label{S-dataselect}

From the flares observed thus far by HEL1OS, events with statistically
significant emission in both CdTe and CZT detectors are selected. CZT bands are background dominated for weaker events, we require each selected flare to have significant emission in at least two CZT energy bands in addition to the CdTe detection. A further selection was required during periods of enhanced particle background. Such intervals occur frequently near solar maximum owing to solar energetic particle (SEP) activity associated with flares and coronal mass ejections. The much longer and generally smoother temporal evolution of SEP enhancements allows flare emission to be distinguished from the particle background in most cases. Flares occurring during gradual SEP enhancements
were therefore retained when the flare profile could be clearly identified,
whereas events coincident with impulsive particle enhancements were excluded.
Application of these criteria resulted in a final sample of 312 flares.

\subsection{Methodology}
\label{sec:S-methodology}

\subsubsection{Statistical Characterisation of the Neupert Effect}
\label{sec:S-neupert_statistics}

The Neupert relationship is first tested statistically across the selected
HEL1OS flare sample,

\begin{equation}
F_{\rm SXR}(t) \propto \int_{t_0}^{t} F_{\rm HXR}(t')\,dt'
\label{eq:neupert}
\end{equation}

where $F_{\rm HXR}$ traces the non-thermal emission and $F_{\rm SXR}$ the
thermal response. For this population-level analysis, CdTe emission below
25~keV is treated as predominantly thermal and CZT emission above 25~keV as
predominantly non-thermal~\citep{Holman2003, McAteer2013-nf}. Correlations are calculated between both individual CdTe bands and cumulative CZT fluence, and between the broadband 10--25~keV CdTe lightcurve and cumulative CZT emission from 25~keV to $E_{\max}$, the
highest energy with significant flare emission above the background.

For each flare, the background-subtracted HXR flux is integrated from the
flare start, $t_0$, to the peak of the thermal emission, $t_{\rm peak}$, and
correlated with the contemporaneous thermal flux. The statistical analysis tests the Neupert relationship at the population level through the correspondence between thermal emission and cumulative HXR fluence. We additionally examine the relationship at the
level of individual flares by asking whether their time-dependent thermal
emission can be reconstructed directly from the observed non-thermal HXR
emission. For this purpose, we construct a forward model in which the HXR
lightcurve provides the energy-release driver and the SXR lightcurve
represents the resulting thermal response. 

\subsubsection{Spectral Separation of Thermal and Non-thermal Emission}
\label{sec:S-spectral_separation}

For the forward modelling, the thermal to non-thermal boundary is determined independently for each event. CdTe PI (pulse-invariant) spectra are accumulated at 60~s cadence. For each integration, the sum of the thermal and non-thermal photon models is folded through the CdTe effective area and spectral redistribution response, and the resulting count spectrum is fitted in detector space (Appendix~\ref{app:eco}). The crossover between the fitted thermal and non-thermal photon components defines $E_{\rm co}(t)$. Since this boundary evolves during a flare, the maximum value

\begin{equation}
E_{\rm co,max} = \max_t \left[E_{\rm co}(t)\right]
\label{eq:ecomax}
\end{equation}

is adopted, with an additional 1~keV tolerance accounting for uncertainty in
the crossover and the detector spectral resolution. The upper boundary,
$E_{\rm high}$, is the highest energy with significant flare emission above the background. For each flare, the background is estimated from a pre-flare interval during which the count rate in the relevant detector and energy band is approximately stationary. This baseline includes instrumental contributions, the cosmic X-ray background, charged-particle-induced counts, and the non-flaring solar emission. Background-subtracted thermal and non-thermal lightcurves are then constructed over 10~keV--$E_{\rm co,max}$ and
$E_{\rm co,max}$--$E_{\rm high}$, respectively. The latter provides the HXR driver for the forward model.

\begin{figure*}[htb!]    
\centerline{\includegraphics[width=0.8\textwidth,clip=]{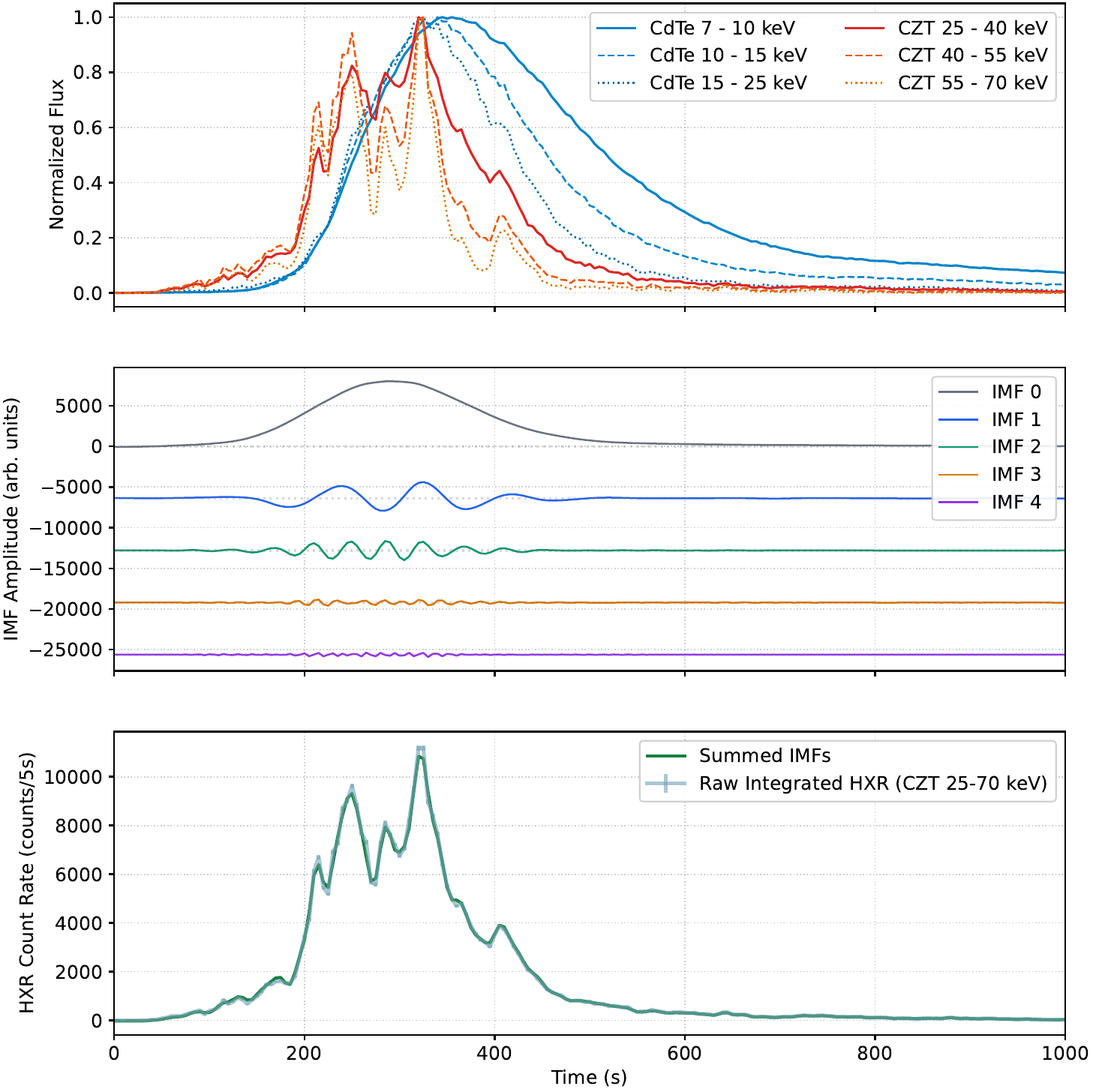}}
\small
        \caption{Temporal structure of the HXR energy-release profile for the flare
SOL2026-04-24T01:00. \textit{Top:} Normalised CdTe and CZT lightcurves,
showing increasingly structured emission at higher energies.
\textit{Middle:} Variational mode decomposition of the 5~s cadence,
band-integrated CZT lightcurve into five intrinsic mode functions
(IMFs). \textit{Bottom:} Reconstruction formed by summing the first four modes while
excluding only the highest-frequency mode, compared with the observed
CZT 25--70~keV lightcurve. This reconstruction illustrates the recovery of
the principal HXR temporal structure by the retained VMD modes.}
\label{vmd_for_flare}
\end{figure*}


\subsubsection{Decomposition of the Non-thermal HXR Driver}
\label{sec:hxr_decomposition}

The non-thermal HXR lightcurve is treated as a superposition of temporally
localised energy-release episodes, consistent with the burst-like structure
commonly observed in flare HXR emission \citep{Aschwanden1998,Collier2024-ts,Collier2023}.
To identify these structures without fitting directly to statistical fluctuations, the background-subtracted HXR lightcurve is first decomposed using Variational Mode Decomposition (VMD)~\citep{Dragomiretskiy2014-fo}. VMD, similar to empirical mode decomposition, separates the signal into a finite set of band-limited intrinsic mode functions (IMFs), thereby isolating variability on different characteristic timescales. Similar to EMD-based approaches, components dominated by high-frequency fluctuations are excluded, and the remaining physically relevant IMFs are summed to obtain a de-noised representation of the HXR profile~\citep{Kolotkov2015-ib,Lim2025-wc,Kanaujiya2026}.

The effectiveness of this decomposition is illustrated in Figure~\ref{vmd_for_flare}, which shows the individual IMFs and the reconstruction obtained by summing the retained modes.
The reconstructed profile recovers the principal temporal structures of the observed 25--70~keV HXR lightcurve, while suppressing the high-frequency fluctuations. This demonstrates
the ability of VMD to separate the dominant HXR variability from noise-dominated components.

For the pulse localisation, the background-subtracted CZT lightcurve is decomposed using five modes (with VMD parameters $\alpha=10$, and $tolerance =10^{-7}$). The intermediate modes, $1-3$, are summed to form the profile used to identify the bursts. The number of modes is fixed empirically at five for this analysis, rather than increased sequentially according to a residual threshold; the intermediate modes retain the burst-scale variations of interest. Candidate peaks are identified in the summed profile using a minimum prominence of 5\% and a minimum height of 3\% of its maximum amplitude. When more than 12 candidates are identified, the 12 most prominent are retained.

The reconstructed VMD profile is used only to \emph{localise} the individual bursts. Peaks in this profile define the initial pulse peak times, $t_i$, while the widths of the corresponding peaks provide initial estimates of their rise and decay widths. The initial pulse amplitude is taken from the original HXR lightcurve at the candidate peak.  These estimates are used to initialise a global fit to the \emph{original, unsmoothed HXR lightcurve}, in which the parameters of all identified pulses are optimised simultaneously.

Each localised burst is represented by an asymmetric Gaussian, allowing the
rise and decay widths to vary independently:

\begin{equation}
\begin{split}
H_i(t) = A_i
\begin{cases}
\displaystyle
\exp\left[-\frac{(t-t_i)^2}{2\sigma_{i,r}^{\,2}}\right],
& \text{for } t \leq t_i, \\[12pt]
\displaystyle
\exp\left[-\frac{(t-t_i)^2}{2\sigma_{i,d}^{\,2}}\right],
& \text{for } t > t_i.
\end{cases}
\end{split}
\label{eq:asymmetric_gaussian}
\end{equation}

Here, $A_i$ is the pulse amplitude, $t_i$ is the time of maximum of the
$i$th pulse, and $\sigma_{i,r}$ and $\sigma_{i,d}$ are its rise and decay
widths, respectively. The global fit optimises these parameters simultaneously
against the measured HXR lightcurve using an unweighted least-squares
objective. The resulting non-thermal driver is

\begin{equation}
H(t) = \sum_{i=1}^{N} H_i(t).
\label{eq:hxr_sum}
\end{equation}

This procedure uses the decomposition to robustly
identify the burst structure while retaining the measured HXR counts for the
final quantitative representation of the energy-release profile. The fitted parameters for the representative flare (shown in Figure~\ref{vmd_for_flare}) are listed in
Table~\ref{tab:hxr_pulse_parameters}.

\begin{deluxetable}{ccccc}
\tablecaption{Fitted parameters of the individual HXR pulses for the representative flare  (SOL2026-04-24T01:00).\label{tab:hxr_pulse_parameters}}
\tablehead{
\colhead{Pulse} & \colhead{$t_i$ (s)} & \colhead{$A_i$ (counts/5 s)} &
\colhead{$\sigma_{i,r}$ (s)} & \colhead{$\sigma_{i,d}$ (s)}
}
\startdata
1 & 113.7 & 621.7 & 16.2 & 3.6 \\
2 & 128.0 & 841.5 & 8.0 & 7.9 \\
3 & 160.0 & 211.7 & 2.2 & 0.9 \\
4 & 170.7 & 1556.0 & 20.5 & 21.7 \\
5 & 214.1 & 4071.0 & 11.9 & 4.0 \\
6 & 248.8 & 9964.0 & 21.7 & 18.8 \\
7 & 284.7 & 6972.0 & 5.4 & 30.6 \\
8 & 322.2 & 8946.0 & 8.6 & 19.2 \\
9 & 367.6 & 3201.0 & 13.4 & 23.9 \\
10 & 408.4 & 2611.0 & 9.1 & 24.5 \\
11 & 470.0 & 628.1 & 38.9 & 59.2 \\
12 & 570.0 & 258.2 & 8.5 & 67.9 \\
\enddata
\end{deluxetable}

\subsubsection{Forward Modelling of the Thermal X-ray Response}
\label{sec:thermal_forward_model}

Each fitted HXR pulse is taken to represent an episode of non-thermal energy
deposition producing a finite thermal response.  Exponential response
functions provide a simple representation of the competition between
impulsive energy input and subsequent plasma losses \citep{Aschwanden2012-cn,2017SoPh..292...77G}. The response to the $i$th
pulse is represented by the causal exponential kernel 

\begin{equation}
K_i(t) =
\begin{cases}
\exp\left(-t/\tau_i\right), & t \geq 0,\\
0, & t < 0,
\end{cases}
\label{eq:cooling_kernel}
\end{equation}

where $\tau_i$ is its effective response timescale. The thermal contribution
from that pulse is

\begin{equation}
S_i(t) = C_i \left[H_i * K_i\right](t)
\label{eq:pulse_convolution}
\end{equation}

where $C_i$ is an amplitude scaling factor and $*$ denotes convolution. The
complete modelled thermal lightcurve is consequently written as,

\begin{equation}
S_{\rm model}(t) =
\sum_{i=1}^{N} C_i
\int_{-\infty}^{t}
H_i(t')\,
\exp\left[-\frac{t-t'}{\tau_i}\right]
dt'.
\label{eq:pulse_convolution_integral}
\end{equation}

Both $C_i$ and $\tau_i$ are allowed to vary between pulses, allowing the
thermal response to evolve between successive energy-release episodes. The fitted
$\tau_i$ is therefore treated as an effective response timescale, incorporating
the unresolved effects of conductive and radiative losses, plasma evolution
and continued heating, rather than as a direct measurement of an individual
cooling process. Finally, $S_{\rm model}(t)$ is fitted to the observed thermal-dominated CdTe
lightcurve. The comparison tests whether the measured thermal evolution can
be represented as the superposition of the responses to the independently
resolved non-thermal HXR bursts.


\begin{figure*}[htb!]    
\includegraphics[width=0.9\textwidth]{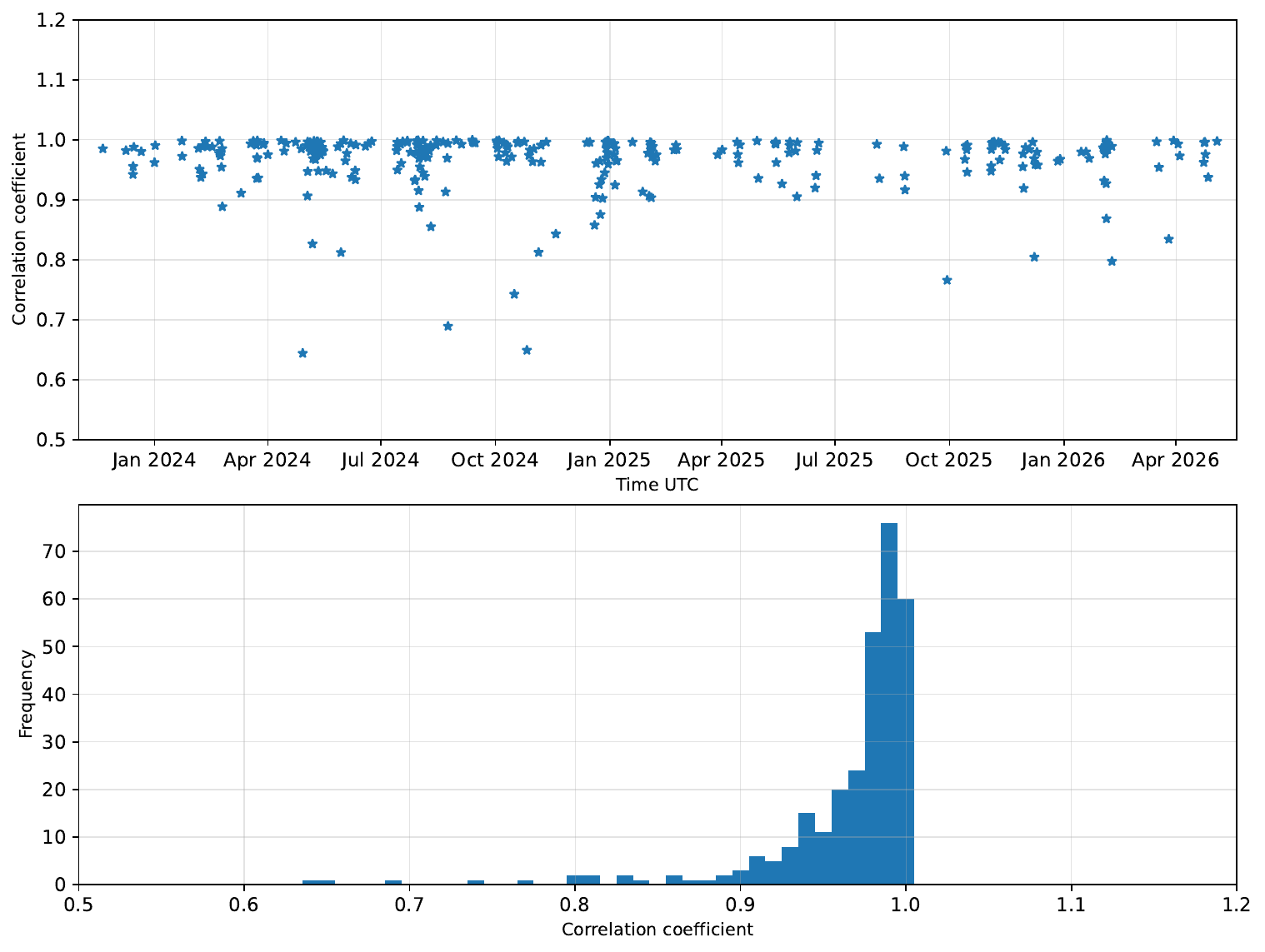}
 \begin{center}
       \caption{Population-level Neupert relationship in the HEL1OS flare sample.
\textit{Top:} Event-wise correlation between thermal-dominated CdTe flux and
cumulative non-thermal CZT fluence over the observing interval.
\textit{Bottom:} Distribution of the correlation coefficients, strongly
concentrated towards unity with a median value of $r=0.984$.}
\label{corr_hist}
 \end{center}

\end{figure*}

\section{Results}
\label{sec:results}

\subsection{Population-level Neupert Relationship}
\label{sec:population_neupert}

We first examine the Neupert relationship across the complete HEL1OS flare
sample. Figure~\ref{corr_hist} shows the correlation between the
thermal-dominated CdTe flux and cumulative non-thermal CZT fluence for each
event. The coefficients remain consistently high throughout the observing
interval, with no evident systematic temporal dependence. Their distribution
is strongly concentrated towards unity, with a median correlation coefficient
of $r=0.984$. 

The relationship remains strong when examined in individual energy bands.
Table~\ref{tab:energy_corr} gives the median coefficients for different
CdTe--CZT combinations. The strongest correspondence is obtained between the
lower-energy bands, with a median coefficient of 0.981 for 7--10~keV CdTe
versus 25--40~keV CZT band. For each thermal band, the correlation
decreases systematically towards higher HXR energies, but remains above
$\sim0.85$ for all combinations considered. The broadband comparison between
10--25~keV thermal emission and cumulative 25~keV--$E_{\max}$ non-thermal
emission gives a median coefficient of 0.984. Together, Figure~\ref{corr_hist}
and Table~\ref{tab:energy_corr} establish Neupert-like evolution as a
population-wide characteristic of the selected HEL1OS flares, while also
revealing a systematic energy dependence in the strength of the relationship.
The population analysis establishes the integral relationship between the
non-thermal and thermal emission. We next examine whether the temporal
structure contained within the HXR emission can account for the observed
evolution of the thermal component.
\begin{table}[ht]
\centering
\caption{Median correlation coefficients between thermal-dominated CdTe flux
and cumulative non-thermal CZT fluence in different energy bands.}
\label{tab:energy_corr}
\begin{tabular}{lccc}
\hline
\hline
CdTe band & \multicolumn{3}{c}{CZT band} \\
\cline{2-4}
 & 25--40~keV & 40--55~keV & 55--70~keV \\
\hline
7--10~keV & 0.981 & 0.953 & 0.876 \\
10--15~keV & 0.973 & 0.953 & 0.867 \\
15--25~keV & 0.958 & 0.940 & 0.855 \\
\hline
\end{tabular}
\end{table}

\subsection{Pulse-resolved Thermal Response}
\label{sec:reconstruction_results}

Figure~\ref{vmd_for_flare} illustrates the temporal structure that underlies the
broadband Neupert relationship. The energy-resolved lightcurves in the upper
panel show a clear progression from the highly structured CZT emission to the
smoother and longer-lived lower-energy CdTe profiles. The principal HXR
bursts occur during the development of the thermal component, while the latter
persists well beyond the strongly impulsive emission.

The VMD components shown in the middle panel separate the HXR variability
over different temporal scales. Their selected combination recovers the
principal structures of the measured HXR lightcurve, as shown in
the lower panel. This representation localises the individual energy-release
episodes while retaining their timing and characteristic temporal structure,
which are subsequently used to initialise the pulse-resolved description of
the measured HXR emission.

\begin{figure*}[htb!]    
\centerline{\includegraphics[trim=0cm 0cm 0cm 0cm,width=1.0\textwidth,clip=]{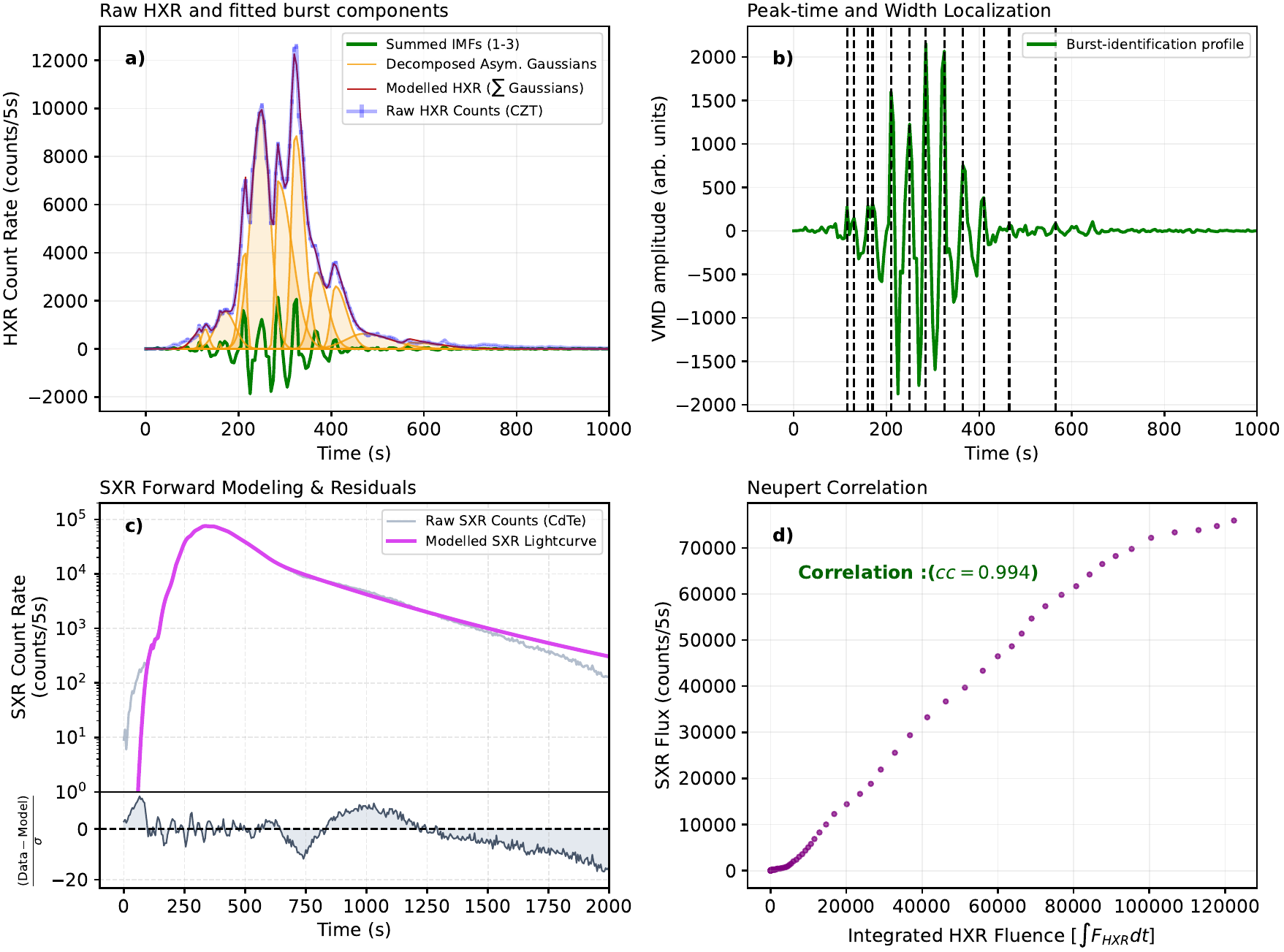}}
\small
      \caption{Pulse-resolved reconstruction of the thermal X-ray response.
\textit{(a)} Measured HXR emission (light blue), VMD-reconstructed profile
(green), individual asymmetric Gaussian components (orange), and their
summed HXR model (red).
\textit{(b)}De-noised VMD profile (green), with dashed vertical lines
marking the peak-times of the localised HXR bursts.
\textit{(c)} Summed modelled thermal response (magenta) compared with the
observed CdTe emission (light blue).
\textit{(d)} Corresponding integral Neupert relation between the thermal
flux and cumulative HXR fluence (purple), demonstrating their close
correspondence.}
\label{hxt_to_sxr_demonstrate}
\end{figure*}

Figure~\ref{hxt_to_sxr_demonstrate} brings together the complete reconstruction for a representative flare. The measured HXR profile is resolved into a sequence of
asymmetric bursts, whose peak-times and
characteristic widths are localised through the VMD-assisted procedure. The resulting summed thermal response
closely follows the observed CdTe lightcurve through the principal rise,
maximum, and much of the extended decay. Thus, the
pulse-resolved treatment and the conventional cumulative description provide
complementary views of the same heating--response relationship: the former
retains the temporal sequence of the energy release that is lost upon
integration.

The reconstruction also exposes departures from this simple correspondence.
The largest differences between the observed and modelled thermal emission
occur preferentially near the beginning of the flare and towards the late
decay, whereas the main flare evolution is reproduced well. They identify phases in which the measured HXR bursts
alone do not fully describe the thermal evolution; their physical
implications are considered in Section~\ref{S-disc}.

\subsection{Response across Different HXR Burst Morphologies}
\label{sec:burst_morphology}

The HXR energy release is not organised in a single characteristic form. A
flare may be dominated by one impulsive burst, contain an irregular sequence
of bursts, or show a more organised, quasi-periodically modulated sequence.
The pulse-resolved framework allows these different temporal morphologies to
be treated within the same description.

\begin{figure*}[htb!]    
\centerline{\includegraphics[width=1.0\textwidth,clip=]{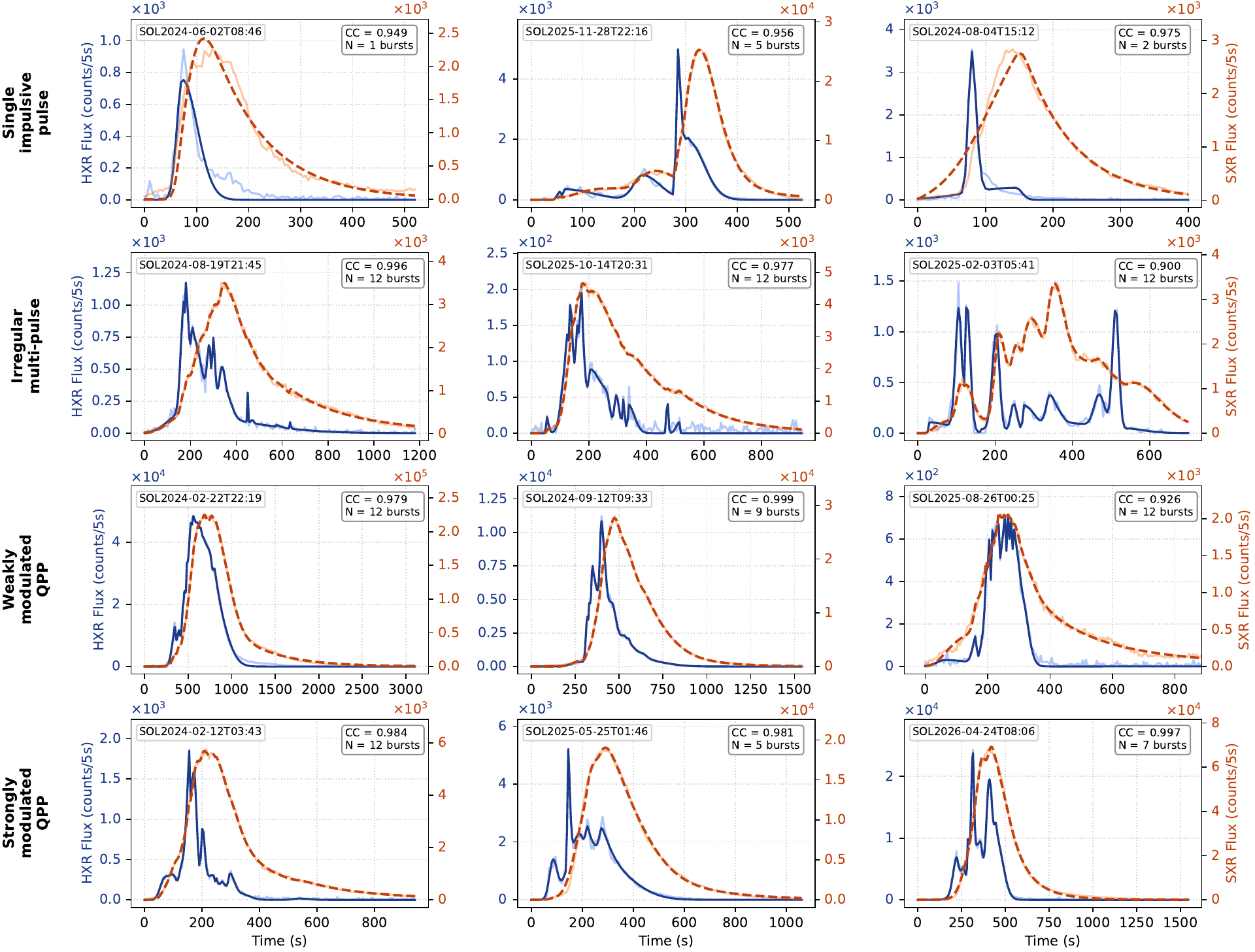}}
\small
        \caption{Thermal-response reconstruction across different HXR burst morphologies. Rows show representative single-impulsive, irregular
multi-pulse, weakly quasi-periodically modulated, and strongly
quasi-periodically modulated events. The legend in the top-left panel
gives the flare start time, while the legend in the top-right panel reports the Neupert correlation coefficient (CC) and the number of bursts identified in the HXR lightcurve.
Light-blue curves show the observed HXR emission, dark-blue curves the
modelled HXR emission, light-orange curves the observed thermal X-ray
emission, and dark-orange dashed curves the modelled thermal response. 
The reconstructions illustrate the performance of the same pulse-response
framework across markedly different temporal organisations of
non-thermal energy release.}
\label{hxr_categories}
\end{figure*}

Figure~\ref{hxr_categories} demonstrates this behaviour for representative
single-impulsive, irregular multi-pulse, weakly quasi-periodically modulated,
and strongly quasi-periodically modulated events. In the single-pulse cases,
a comparatively short HXR energy-release episode produces a substantially
broader thermal response. For irregular multi-pulse events, several distinct
HXR bursts contribute to a common thermal envelope, and the individual
responses increasingly overlap as the burst sequence becomes more complex.
Despite these markedly different non-thermal profiles, the observed thermal
evolution is reproduced by the same pulse-response formalism.

The lower two rows of Figure~\ref{hxr_categories} extend this comparison to
events whose HXR peaks show a more organised temporal modulation. Such
pulse-timing patterns are commonly used to characterise fast HXR
variability and candidate quasi-periodic behaviour
\citep{Collier2023}. Here, however, the categories are morphological and
do not constitute formal QPP detections. In the weakly modulated cases, the
short-timescale HXR structure is largely absorbed into a smooth thermal
envelope. When the modulation is stronger, the correspondence between
successive HXR energy-release episodes and structure in the thermal evolution
becomes more apparent. In both cases, however, the thermal response remains
substantially smoother than the underlying non-thermal driver.

\section{Discussion}
\label{S-disc}

The present analysis places the classical Neupert effect in a more explicit
spectral context. Its usual formulation as an HXR--SXR relationship implicitly
associates the two energy ranges with non-thermal and thermal emission,
respectively. The energy-resolved HEL1OS analysis shows that the more relevant
distinction is indeed between these two emission components rather than between
fixed photon-energy bands. The Neupert correspondence remains strong across
different band combinations provided that the lower-energy emission is
predominantly thermal and the higher-energy emission predominantly
non-thermal.  This is consistent with the thick-target interpretation
\citep{Brown1971-jh,2011SSRv..159..107H}. The decrease in the extent of correlation towards higher HXR energies may reflect the smaller energetic contribution of the high-energy electron population, although decreasing counting statistics and the increasing difficulty of separating weak flare emission from background at these energies may also contribute.

This conclusion is strengthened by comparison with independent statistical
studies. \citet{Veronig2002-ms} found strong Neupert behaviour among a large
BATSE--GOES sample, although with a substantial population departing from the
simplest timing expectation. More recently, \citet{2024SoPh..299...57L} found Neupert correlations above 0.9 for all flares used in their study.
The latter observations were obtained with different instruments and during
an earlier interval of Solar Cycle 25 than the HEL1OS sample. The present
results therefore extend this behaviour over an independent flare population
and, through the energy-resolved analysis, emphasise that the underlying
relationship is naturally regarded as one between non-thermal energy
release and the ensuing thermal response, rather than simply between HXR and
SXR photon bands.

\subsection{What the Departures from the Reconstruction Represent}
\label{sec:residuals}

The close overall reconstruction should not imply that either the thermal or non-thermal component can be perfectly isolated spectrally. Thermal bremsstrahlung extends into the HXR range, while non-thermal bremsstrahlung extends below the spectral crossover energy. The crossover itself depends on the evolving thermal and accelerated-electron populations
\citep{1990SoPh..129..113L,2011SSRv..159..107H,Aschwanden2019-su}. Our use of an event-dependent $E_{\rm co}$ therefore minimises, but cannot eliminate, this spectral overlap.

This is particularly relevant during strongly impulsive intervals. A rapid
increase in the accelerated-electron population can produce non-thermal
variability extending into energies otherwise dominated by thermal emission.
Consequently, some short-timescale structure in the nominal thermal
lightcurve may be contemporaneous non-thermal emission rather than the
hydrodynamic response of the heated plasma. Such spectral mixing provides a
natural source of local discrepancies between the reconstructed and observed
lightcurves and places a limit on how literally individual small residuals
should be interpreted.

A separate limitation arises when the flare heating is represented by a single impulsive episode. \citet{Qiu2021-yk} tested the same class of empirical convolution model using the HXR light curve itself as the heating proxy, corresponding to \(N=1\), and found that this formulation could not reproduce the observed SXR evolution satisfactorily. 
A similar behaviour is evident in our results. The first row of Figure 4, corresponding to a predominantly single-burst HXR profile, gives the least consistent reconstruction of the observed thermal X-ray emission among the representative cases. The HXR pulse produces a broad thermal response, but parts of the observed thermal evolution are not accounted for by that single response. In contrast, when the HXR emission contains several temporally distinct bursts, the overlapping responses provide a closer representation of the thermal envelope.

The temporal profiles of individual bursts also merit comparison with the rapid-rise, exponential-decay morphology reported in solar radio observations. \citet{1992ApJ...384..656W} described an impulsive burst whose time profiles were closely similar at 8.6, 15, and 86~GHz. Related spiky bursts at 17 and 34~GHz, with rise times of a few seconds followed by rapid exponential decay, were subsequently reported by \citet{2001ApJ...547.1090K}. This morphology may not, however, be generalised to HXR emission, as  \citet{2003dysu.book.....D} note that the rapid-rise, exponential-decay profile seen in some millimetre bursts is not generally shared by the corresponding 25--100~keV HXR profiles, which can have substantially different temporal shapes. The radio results therefore provide a comparison of impulsive time-profile phenomenology, but do not imply that HXR bursts follow the same profile. 

The more systematic departures at the beginning and end of the flare carry
greater physical significance. An early thermal excess is consistent with observations of
pre-flare or ``hot-onset'' plasma preceding detectable HXR emission
\citep{2021MNRAS.501.1273H}. Importantly, this does not by itself demonstrate
a non-beam heating mechanism. The energy content of a flare electron
distribution is extremely sensitive to its low-energy extent
\citep{2011SSRv..159..107H,Aschwanden2019-su,Kontar2015-zq}. Electrons below the
non-thermal band accessible with sufficient HEL1OS statistics may therefore
deposit appreciable energy while producing little detectable emission in the
HXR driver used here. Effective area limitations at these energies further
restricts the ability to separate a weak non-thermal component from the much
stronger thermal continuum. The early positive residual should consequently
be interpreted more generally as heating that is not represented by the
\emph{measured} non-thermal HXR driver. Possible contributions include
low-energy accelerated electrons, direct coronal heating and conductive
energy transport.

The late decay presents a different problem. Here the observed thermal
emission can persist above the response predicted from the impulsive HXR
bursts. Then, a purely passive response to the measured impulsive heating is
insufficient. \citet{Veronig2002-ms} found that approximately one quarter of their flare sample showed continued SXR growth after the termination of detectable HXR emission,
which they interpreted as evidence for an additional energy-transport or
heating component. From the cooling of 72 M- and X-class flares,
\citet{Ryan2013-oq} likewise found observed cooling times systematically longer
than predicted by an analytical model without continued heating. Thus, the
late-phase residual in the present reconstruction is not merely a fitting
limitation: at least in some events, it is consistent with prolonged energy
input after the dominant non-thermal impulsive phase. An extended SXR decay can
result from continued heating, from a succession of independently cooling
loops, or from plasma structures with different conductive and radiative
timescales. An unresolved X-ray lightcurve alone cannot distinguish these possibilities. The residual therefore identifies \emph{when} the HXR-constrained heating history becomes insufficient, while spatially resolved observations are required to establish \emph{why}.

\subsection{Pulse-dependent Response as a Diagnostic of Flare Evolution}
\label{sec:response_diagnostics}

A further consequence of resolving the HXR driver into individual bursts is
that a single thermal response need not be imposed on the complete flare.
The fitted $C_i$ and $\tau_i$ instead characterise how the response associated
with successive energy-release episodes changes as the flare evolves. Their
physical interpretation must be made cautiously. In particular, $\tau_i$ is
an effective response timescale rather than a direct measurement of either
the conductive or radiative cooling time. Conductive cooling depends strongly
on temperature and loop length, while radiative cooling depends on the plasma
density and radiative loss function; density evolution, enthalpy exchange and
continued heating further couple these processes
\citep{1995ApJ...439.1034C,2008ApJ...682.1351K,2012ApJ...752..161C}.

Nevertheless, the \emph{relative evolution} of the fitted parameters can be
diagnostic. A sequence of bursts with comparable response times and scaling
factors would be compatible with repeated energisation of structures having
similar thermodynamic properties. Conversely, a systematic change or large
pulse-to-pulse dispersion indicates that the flare cannot be represented by
repeated heating of a stationary plasma response. One plausible scenario is
the progressive reconnection of an arcade: successive energy-release episodes
energise different loops with changing lengths, volumes and pre-event
densities. Longer loops, for example, generally introduce longer conductive
timescales, whereas changes in density alter both emission measure and
radiative cooling. Variations in $C_i$ can likewise reflect changes in the
amount of plasma contributing to the observed thermal emission for a given
HXR pulse. The temporal parameters alone cannot establish this topology, and therefore we do not identify changes in $\tau_i$ uniquely with changing loop
length. Instead, their value is as a rapid unresolved diagnostic that
generates a testable prediction for imaging observations.

\subsection{Connection to Hydrodynamic and UV/EUV Flare Modelling}
\label{sec:hydrodynamic_connection}

The pulse-resolved HXR profile, like the example shown in Figure~\ref{hxt_to_sxr_demonstrate}, also provides a practical bridge between observations and hydrodynamic modelling. Zero-dimensional models such as EBTEL \citep{2008ApJ...682.1351K,2012ApJ...752..161C} calculate the loop-averaged thermodynamic evolution of a coronal loop for a specified heating function and are particularly suited to modelling large ensembles of independently heated strands. A persistent challenge in forward hydrodynamic modelling is that the temporal form of the heating function must generally be specified. Here, the timings, durations and relative strengths of the resolved HXR bursts provide an observation-driven constraint on that input.
Rather than assuming identical triangular or Gaussian heating episodes, the resolved HXR burst sequence can therefore provide an observation-driven constraint on the temporal structure of the heating function used in future hydrodynamic calculations. To convert the HXR sequence into a physical heating function, however, the observed HXR emission would need to be combined with spectral constraints on the accelerated-electron distribution and, where required, the source geometry. Such a physically calibrated heating function could then be supplied to an EBTEL or hydrodynamic calculation to determine the temperature and density evolution and, with the appropriate loop geometry and emissivity, the resulting emission measure and synthetic X-ray emission \citep{Costa2016}.

An especially useful comparison is with the UV-based modified Neupert
framework of \citet{Qiu2021-yk}. In that approach, spatially resolved UV
footpoint lightcurves are used as proxies for heating in individual flare
loops; each is convolved with a cooling kernel and the loop contributions are
summed to reproduce the observed SXR emission. Their modelling showed that a
collection of purely impulsive heating events was insufficient in some
events, while an additional gradual heating component substantially improved
the SXR reconstruction. This result is directly relevant to the late-phase
departures found here.

The two approaches constrain different parts of the same energy-transport
problem. UV footpoint emission provides spatial localisation of newly
energised magnetic structures, while HXR emission diagnoses the accelerated-electron population, although the conversion from photon flux to deposited energy remains sensitive to the electron spectrum, particularly its low-energy cutoff. The present
HXR-driven model therefore supplies strong temporal information on the
non-thermal energy-release episodes, while UV/EUV observations supply the
spatial and thermodynamic context absent from Sun-as-a-star X-ray
lightcurves.

A combined HXR--UV--EUV--SXR treatment is consequently a natural extension
of this work. Individual HXR bursts could be associated with newly brightened
UV footpoints and subsequently formed EUV loops, while the corresponding SXR
response constrains their integrated thermal evolution. Such an analysis
could determine whether changes in $C_i$ and $\tau_i$ accompany the
progressive formation of an arcade, distinguish prolonged heating from the
cooling of successively formed loops, and test whether the thermal energy
unaccounted for by the HXR reconstruction has a spatially identifiable
counterpart. The same HXR burst sequence could then be used as the heating
constraint for multi-strand hydrodynamic modelling. The Neupert relationship
would thereby become not simply a correspondence between two lightcurves, but
an observational framework linking particle acceleration, energy deposition,
magnetic restructuring and the subsequent thermodynamic evolution of the
flaring plasma.

\section{Conclusion}
\label{S-conclusion}

Using HEL1OS observations, we show that the Neupert relationship is a robust
population-level property of flares with detectable non-thermal emission and
extend this relationship beyond its conventional integral form. By resolving
the HXR emission into individual energy-release episodes, the observed thermal
X-ray evolution can be reconstructed as the superposition of their plasma
responses, irrespective of whether the HXR driver consists of isolated,
irregular, or quasi-periodically organised bursts. The departures from this reconstruction are themselves informative, with the early and late phases identifying thermal evolution not fully represented by the measured HXR driver. The pulse-resolved Neupert framework therefore
provides an observation-driven link between particle acceleration, energy
deposition, and the subsequent plasma response. Combined with spatially
resolved UV/EUV observations and hydrodynamic modelling, it offers a route
towards connecting individual episodes of non-thermal energy release to the
evolving magnetic and thermodynamic structure of solar flares.


\begin{acknowledgments}
Aditya-L1 is an observatory-class mission fully funded and operated by the Indian Space Research Organisation (ISRO). The mission was conceived and realised with contributions from various ISRO centres. The science payloads and science-ready data products were realised by the respective payload PI institutes in close collaboration with ISRO centres. We acknowledge the use of HEL1OS data from the Aditya-L1 mission, archived at the Indian Space Science Data Centre (ISSDC). We also acknowledge the contributions of the URSC and ISTRAC departments that supported the development and operation of the instrument. SPT and AKS thank IIT-BHU for supporting this work. The authors from ISRO thank the Deputy Director, Payload Data Management and Space Astronomy Area (PDMSA), the Associate Director, and the Director of URSC for their encouragement and continued support. The authors are grateful to the anonymous referee whose comments improved the manuscript.
\end{acknowledgments}

\section*{Additional statements}

The authors declare that they have no conflicts of interest with regard to this work.

\clearpage
\appendix
\restartappendixnumbering
\section{Estimation of the Thermal to Non-thermal Crossover Energy}
\label{app:eco}

The thermal--non-thermal crossover energy, $E_{\rm co}$, is defined as the
photon energy at which the fitted thermal and non-thermal components have equal flux. For each selected flare, the observed count spectrum is fit with the sum of the thermal model \texttt{f\_vth} and the thick-target
bremsstrahlung model \texttt{thick\_fn}. The combined photon model is folded
through the CdTe effective area and spectral redistribution response, and the
fit is performed in detector-count space.

\begin{figure}[!htbp]
  \centering
  \includegraphics[width=0.70\textwidth]{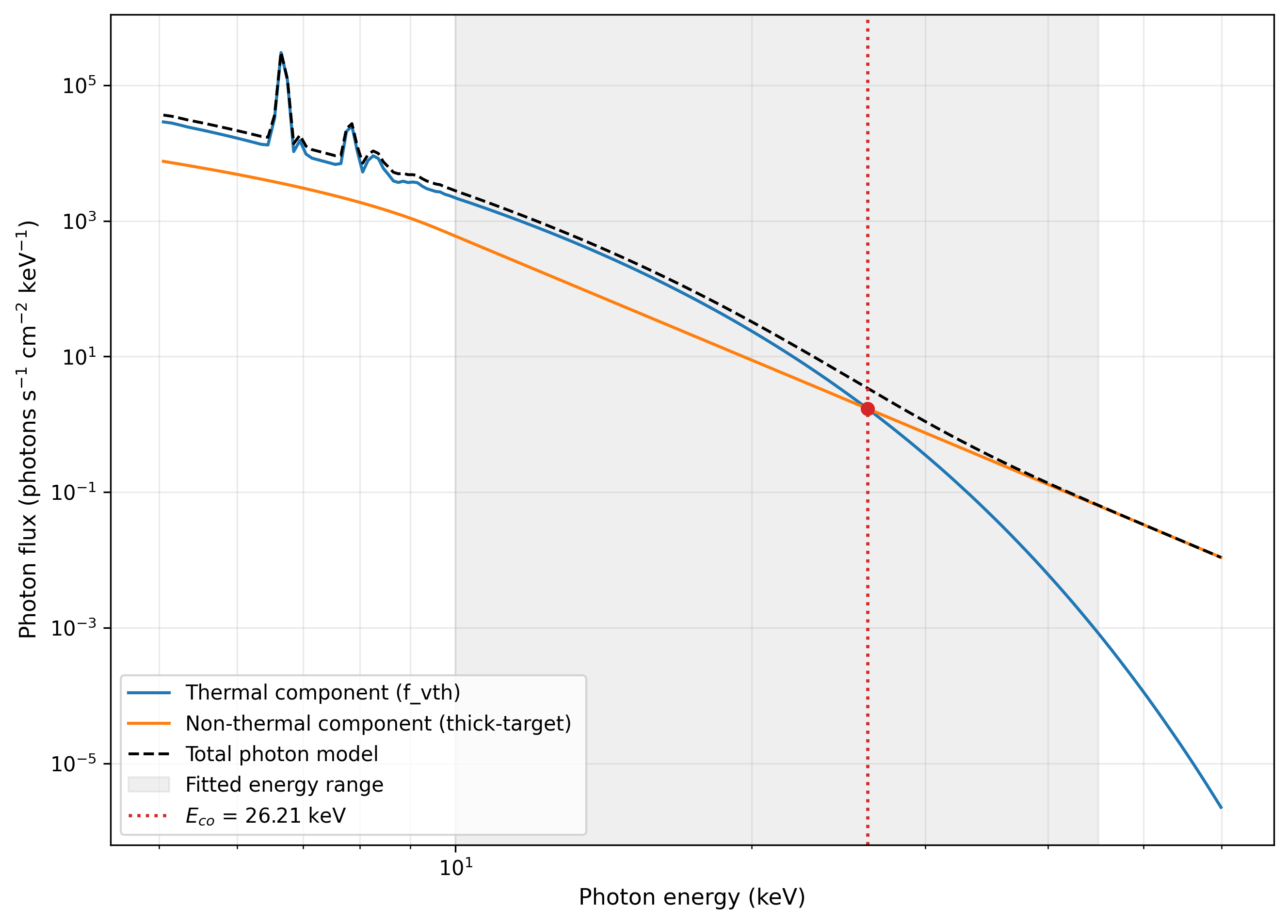}
  \caption{Example of the thermal--nonthermal decomposition in photon space
  for a CdTe flare spectrum. The solid blue and orange curves show the fitted
  thermal ($\texttt{f\_vth}$) and nonthermal
  ($\texttt{thick\_fn}$) photon components, respectively; the dashed  black curve
  shows their sum. The shaded region indicates the energy interval used in
  the fit. The vertical dotted line marks their intersection,
  $E_{\rm co}=26.21$~keV.}
  \label{fig:eco_example}
\end{figure}

After fitting, the two components are evaluated separately on the incident
photon-energy grid, before response folding. The crossover is obtained from
\begin{equation}
  I_{\rm th}(E_{\rm co}) = I_{\rm nth}(E_{\rm co}),
  \label{eq:eco_definition}
\end{equation}
where $I_{\rm th}$ and $I_{\rm nth}$ denote the thermal and non-thermal photon
spectra, respectively. This procedure is applied to each 60~s CdTe integration
for which a reliable fit is obtained, yielding a time-dependent crossover
$E_{\rm co}(t)$. The flare boundary used in the analysis is the maximum
crossover, $E_{\rm co,max}=\max_t E_{\rm co}(t)$. A crossover is retained
only when the intersection lies within the energy interval used in the fit;
an intersection outside this interval would be an extrapolation and is not
treated as a directly constrained estimate. The photon crossover energy is
distinct from the low-energy cutoff, $E_{\rm c}$, of the non-thermal electron
distribution. Figure~\ref{fig:eco_example} shows an example of the
decomposition. 

\bibliographystyle{aasjournal}
\bibliography{references}  


\end{document}